\documentclass[11pt]{article}
\usepackage{amsmath,amssymb}
\usepackage{graphicx}
 \allowdisplaybreaks
 \newcommand{\be}{\begin{equation}}
 \newcommand{\ee}{\end{equation}}
 \newcommand{\bea}{\begin{eqnarray}}
 \newcommand{\eea}{\end{eqnarray}}
 
 \newcommand{\nn}{\nonumber}
 \newcommand{\td}{\tilde}
 \newcommand{\wtd}{\widetilde}
 \newcommand{\pd}{\partial}

 \newcommand{\bm}{{\bf m}}
 \newcommand{\bn}{{\bf n}}

 \newcommand{\cL}{{\cal L}}

 \newcommand{\cO}{{\cal O}}

 \newcommand{\cV}{{\cal V}}
 
 \newcommand{\tda}{{\td a}}
 \newcommand{\tdb}{{\td b}}

\long\def\symbolfootnote[#1]#2{\begingroup%
\def\thefootnote{\fnsymbol{footnote}}\footnote[#1]{#2}\endgroup}

\newcommand{\aei}{\it Max Planck Institute for Gravitational Physics
(Albert Einstein Institute)\\ Am M\"uhlenberg 1, 14476
Potsdam-Golm, Germany}

\begin{document}
\thispagestyle{empty}
\begin{flushright}
\hfill{AEI-2013-257}
\end{flushright}
\begin{center}

~\vspace{20pt}

{\Large\bf Emergent Symmetry on Black Hole Horizons}

\vspace{25pt}

Jianwei Mei\symbolfootnote[1]{Email:~\sf jwmei@aei.mpg.de}

\vspace{10pt}\aei

\vspace{2cm}

\centerline{\bf Abstract}
\end{center}

For a stationary and axisymmetric black hole, there is a natural
way to split the fields into a probe sector and a background
sector. The equations of motion for the probe sector enjoy a
significantly enhanced symmetry on the black hole horizon. The
extended symmetry is conformal in four dimensions, while in higher
dimensions it is much bigger. This puts conformal symmetry at the
bottom of the ladder of symmetries that can arise on black hole
horizons in generic dimensions.

 \newpage


A black hole is characterized by the presence of a horizon. So
symmetries attached to the horizon presumably hold the key to the
mysteries of black holes. There have been good evidence for this
in relation to the black hole entropy.

For extremal black holes, one can zoom in the near horizon region
and find evidence that quantum gravity in this background is dual
to a conformal field theory in a state having the same
Bekenstein-Hawking entropy as the black hole (see, e.g.
\cite{Strominger:1997eq,Guica:2008mu}). For non-extremal black
holes, one can similarly calculate the black hole entropy through
a putative dual conformal field theory by imposing certain
boundary conditions on (near) the horizon and finding conformal
symmetries there (see, e.g.
\cite{Carlip:1998wz,Carlip:2002be,Carlip:2011ax,Mei:2012wd}).

All these works rely on imposing extra conditions on a chosen
boundary, but sometimes one can find more than one working
boundary conditions for a given background. On the other hand, the
dynamics governing the properties of a black hole is unique. This
raises the question of whether external information like an
artificial boundary condition is necessary in calculating the
black hole entropy.

In this respect, we have found previously that some conformal
symmetries actually emerge on the black hole horizons by
themselves \cite{Mei:2011az}. (See also
\cite{Carlip:2002be,Castro:2010fd}.) The construction only depends
on a few very general properties needed for the metric to describe
a black hole\footnote{The analysis only depends on the general
metric (\ref{metric.general}), which covers other interesting
objects such as black rings \cite{Emparan:2001wn}. But we will
refer to all of them as black holes just for simplicity.}, but
requires no extra boundary conditions. Since the radial direction
is interpreted as the renormalization scale, such symmetries can
be viewed as emergent from the duality perspective. We emphasize
that the new transformations are not the usual diffeomorphisms,
but are a type of gauge transformations of the fields.

The purpose of this paper is to show that, in higher dimensions
($D>4$), the emergent symmetry group is in fact much larger than
that of the conformal symmetries. As a proof of principle, we will
only present the result for the simplest nontrivial case.
Comparing to four dimensions, gravity in higher dimensions is
known to display much richer features, such as more possibilities
for the topology of a horizon \cite{Emparan:2001wn}. What we find
here adds another potentially useful piece to the category.

Our starting point is the metric for a stationary and axisymmetric
black hole,
\be ds^2=f\Big[-\frac{\Delta}{v^2}dt^2+\frac{dr^2}\Delta\Big]
+h_{ij}d\theta^i d\theta^j+g_{ab}(d\phi^a-w^adt)(d\phi^b-w^bdt)
\,,\label{metric.general}\ee
where the details are explained in \cite{Mei:2011az}. Here we only
note that $\phi^a$ and $t$ are ignorable coordinates: $f, v,
h_{ij}, g_{ab}$ and $w^a$ only depend on $r$ and $\theta^i$, while
$\Delta=\Delta(r)$; the horizon is located at $r_0$ with
$\Delta(r_0)=0$, where all other functions are regular. One can
write (\ref{metric.general}) as $ds^2=\wtd{G}_{\mu\nu}dX^\mu
dX^\nu=H_{IJ}dx^Idx^J+G_{AB} dy^A dy^B$, where $x^I\in\{r,
\theta^i\}$ and $y^A\in\{\phi^a,t\}$. Reducing the
Einstein-Hilbert action on the ignorable coordinates $\phi^a$ and
$t$, we find
\bea S&=&\int d^{\td{k}}xd^ky\sqrt{-\wtd{G}}\;\Big(\wtd{R}
-2\Lambda\Big)=(2\pi)^{k-1}T\int d^{\td{k}}x\,\cL\,,\nn\\
\cL&=&\sqrt{|H|}\sqrt{|G|}\;\Big\{R-2\Lambda+\frac14\pd G_{AB}\pd
G^{AB}+(\pd\ln\sqrt{|G|})^2-\frac2{\sqrt{|G|}}\nabla^2\sqrt{|G|}
\Big\}\nn\\
&=&\sqrt{H g/\varrho}\,\Bigg\{R-2\Lambda+\frac14\pd
g_{ab}\pd g^{ab}+\frac\varrho2 g_{ab}\pd w^a\pd w^b\nn\\
&&\qquad\qquad+(\pd\ln\sqrt{g/\varrho}\,)^2
-(\pd\ln\sqrt\varrho\,)^2-\frac{2\nabla^2\sqrt{g/\varrho}
}{\sqrt{g/\varrho}}\,\Bigg\}\,,\label{action.BH}\eea
where $\td{k}=D-k$, $k=[\frac{D+1}2]$ is the total number of
ignorable coordinates, $T$ is the total age of the system,
$\wtd{R}$ and $R$ are the Ricci scalar for $\wtd{G}_{\mu\nu}$ and
$H_{IJ}$, respectively, $H=\det|H_{IJ}|=\frac{f}\Delta
\det|h_{ij}|$, $G=\det|G_{AB}|=-g/\varrho$, $g=\det|g_{ab}|$, and
$\varrho=\frac{v^2}{f \Delta}$. The indices from $H_{IJ}$ have
been suppressed in (\ref{action.BH}).

The basic idea is to treat (\ref{action.BH}) as an action for the
fields $G_{AB}$ living in the fixed background of $H_{IJ}$.
Without considering the back reaction on $H_{IJ}$, there is more
possibility for $G_{AB}$ than in the full theory. The extended
transformations found below will leave the equations of $G_{AB}$,
i.e. $\delta S/\delta G_{AB}=0$, invariant on the horizon, but
will not for $\delta S/\delta H_{IJ}=0$. Although this may appear
{\it ad hoc}, it is not too much different from introducing an
external field to probe the black hole background
\cite{Castro:2010fd}. In the present case, $H_{IJ}$ is playing the
role of a background, while $G_{AB}$ is playing the role of a
probe. As it happens, one relies on the hope that the extended
transformations could become a symmetry of the full system when
the Einstein-Hilbert action is replaced by that of a complete
theory of quantum gravity.

We will focus on the equations of $G_{AB}$ from now on. Varying
$G_{AB}$, we find
\bea\delta\cL&=&\sqrt{H}\Big[E_{ab}\delta g^{ab}+E_a\delta w^a
+E_{(g/\varrho)}\delta\sqrt{g/\varrho}+\nabla_IJ_\delta^I\Big]
\,,\label{deltaL}\\
J_\delta^I&=&\pi^{Iab}\delta g_{ab}+\pi^I_a\delta w^a +\pi^I_{
(g/\varrho)}\delta\sqrt{g/\varrho}\,-2\pd^I\delta
\sqrt{g/\varrho}\,,\label{J.def}\\
\pi^{Iab}&=&\sqrt{g/\varrho}\Big(\frac12\pd^Ig^{ab} -\pd^I
\ln\sqrt\varrho\,g^{ab}\Big)\,,\nn\\
\pi^I_a&=&\sqrt{g/\varrho}\,\varrho g_{ab}\pd^I w^b\,,\quad
\pi^I_{(g/\varrho)}=2\pd^I\ln\sqrt{g}\,.\nn\eea
where the index $I$ goes over $r$ and $\theta^i$. One can firstly
read off the equations of motion $E_{ab}=E_a=E_{(g/\varrho)}=0$,
which are equivalent to (the indices $A,B$ run over $\phi^a$ and
$t$)
\be E_{AB}=-\frac12\frac{\nabla(\sqrt{-G}\,\pd G_{AB})}{\sqrt{-G}}
+\frac12\pd G_{AC}\pd G_{BD}G^{CD}-\frac{2\Lambda}{D-2}G_{AB}
\propto\frac{\delta S}{\delta G^{AB}}=0 \,.\label{eom.G}\ee
All the extended symmetries that we will find satisfy
\be\sqrt{H}\,\nabla_I J_\delta^I \sim\delta E_{AB}
\sim\cO(\Delta)\,.\label{covJ.def}\ee

Our calculation will rely on the following universal property
\be w^a(r,\theta^i)=w^a_0(r)+w^a_1(r,\theta^i)\Delta+\cdots\,,
\label{prop.w}\ee
where ``$\cdots$" stands for higher order terms in the near
horizon limit. This property is necessary for any black hole in
(\ref{metric.general}) to have an intrinsically regular horizon.
Our construction will crucially depends on $w^a\neq0$. In a
coordinate system that is static at the spatial infinity (with
$\phi^a$'s normalized to be of period $2\pi$), $w_0^a(r_0)
=\Omega^a$ is the constant angular velocity of the horizon along
$\phi^a$. For a Schwarzschild like black hole without an intrinsic
rotation, one can firstly change to a rotating frame and then the
same construction can still be applied.

Due to the freedom in redefining the ignorable coordinates, the
action (\ref{action.BH}) has a rigid $GL(k,R)$ symmetry, $G_{AB}\;
\to\;(\cV\cdot G\cdot \cV^T)_{AB}$, where $\cV$ is a $k\times k$
constant matrix. The scaling factor in $GL(k,R)$ is orthogonal to
all other transformations and will not be needed here. We will
only focus on the $SL(k,R)$ subsector in this
work.\footnote{Consequently, one has $\delta \sqrt{-G}\,
=\delta\sqrt{g/\varrho} =0$, which is then assumed to be true for
all the extended symmetries. Since this assumption is made in the
whole spacetime, it is not a boundary condition that we set out to
avoid. When this assumption is dropped, one will be looking at an
extension involving the constant scaling factor of $GL(k,R)$. This
possibility has not been considered carefully.}

In \cite{Mei:2011az} it has been found that some $SL(2,R)$
sub-sector of this $SL(k,R)$ can be extended, on the black hole
horizon, to a centerless Virasoro algebra (Witt algebra),
\be[\delta_m\,,\,\delta_n]=(m -n)\delta_{m +n}\,,\quad
m,n=0,\pm1,\pm2,\cdots\,. \label{algebra.witt}\ee
To do this, one of the rotations (say $\phi^1$) is singled out,
labelled simply as $\phi$, while all other rotations are indexed
with $\td{a}$, $\td{b}$, $\cdots$, i.e.,
\be (G_{AB})=\left(\begin{matrix}g_{\phi\phi}&g_{\phi\tdb}
&-w_\phi\cr g_{\tda\phi}&g_{\tda\tdb}&-w_\tda \cr-w_\phi
&-w_\tdb&-\frac1\varrho+w^2\end{matrix}\right)\,.\ee
Then one considers the $SL(2,R)$ generators, which are $k\times k$
matrices, with only the following non-trivial elements,
\be (L_1)_{1k}=-2(L_0)_{11}=2(L_0)_{kk}=-(L_{-1})_{k1}=1\,.
\label{sl2r.g2}\ee
Using $\delta G=-(L\cdot G+G\cdot L^T)$, one can find how
$\delta_0$ and $\delta_{\pm1}$ act on the fields of $G_{AB}$. Then
the extension to other generators is constructed. Technically,
only $\delta_{\pm2}$ were given explicitly in \cite{Mei:2011az},
which is enough to reconstruct the whole algebra. Here we write
down the general result,\footnote{This result differs from that in
\cite{Mei:2011az} at the sub-leading order. The result here
satisfies the Witt algebra (\ref{algebra.witt}) at the leading
order and also obeys (\ref{covJ.def}); while those in
\cite{Mei:2011az} satisfies the Witt algebra up to the subleading
order, but has $\delta E_{AB} \sim\Delta'(r_0)+\cO(\Delta)$. }
\bea\delta_m g_{\phi\phi}&=&-(m+1)\Big[(m-1) g_{\phi \phi}-m
\frac{w_\phi}{w^\phi} +m(m-1)\frac{w'^\phi(w_\phi-g_{\phi\phi}
w^\phi)}{(w^\phi)^2\Delta'/\Delta}\Big](w^\phi)^m\,,\nn\\
\delta_m g_{\tda\phi}&=&-\frac{m+1}2\Big[(m-1)g_{\tda\phi}
-m\frac{w_\tda}{w^\phi}+m(m-1)\frac{w'^\phi(w_\tda-g_{\tda
\phi}w^\phi)}{(w^\phi)^2\Delta'/\Delta}\Big](w^\phi)^m\,,\nn\\
\delta_m g_{\tda\tdb}&=&0\,,\quad\Longrightarrow\quad
\delta_m\varrho=\varrho\, (m+1)(w^\phi)^m\,,\nn\\
\delta_m w^\phi&=&-\Big[w^\phi+\frac{m(m+1)}2
\frac{g^{\phi\phi}}{\varrho\,w^\phi}\Big](w^\phi)^m\,,\nn\\
\delta_m w^\tda&=&-\frac{m+1}2\Big[w^\tda +m\frac{g^{\tda
\phi}}{\varrho\,w^\phi}\Big](w^\phi)^m\,,\label{delta.m}\eea
where a `prime' means a derivative with respect to $r$. Note the
transformation of $\varrho$ is always determined by $\delta\sqrt{
g/\varrho}=0\;\Rightarrow\;\delta\varrho=\varrho g^{ab}\delta
g_{ab}$.

The subleading terms (those containing a factor $1/\varrho\sim
\Delta$) are important for the symmetry to work. Without them, one
would have found
\be\sqrt{H}\,\nabla_I J_\delta^I \sim\delta E_{AB} \sim
\Delta'(r_0)+\cO(\Delta)\,.\ee
But for extremal black holes, which has $\Delta'(r_0)\sim T_H=0$,
where $T_H$ is the Hawking temperature of the black hole, the
symmetry works without the need for the subleading terms.
Technically this will bring a significant simplification over the
calculation when one tries to make the extension of the full
$SL(k,R)$ algebra.\footnote{The possibility of an extension of the
full $SL(k,R)$ was first suggested to me by Evgeny Skvortsov.}

Once there is an extension to the $SL(2,R)$ sub-sector, the
dressing up of the rest of the $SL(k,R)$ generators is
straightforward. One can simply do this by taking commutators of
the other $SL(k,R)$ generators with (\ref{delta.m}) repeatedly. In
four dimensions, $k=2$ and the conformal symmetry is all that is
there. The simplest non-trivial case is that of the five
dimensional extremal black holes.

In five dimensions, $k=3$ and we will have the following fields to
consider: $g_{11}$, $g_{12}$, $g_{22}$, $w^1$ and $w^2$. As said
above, one can drop the subleading $(1/\varrho\sim \Delta)$-terms
for extremal black holes. After some effort, we find that all the
symmetry generators organize into the following two,
$\delta^+_{m,s}$ and $\delta^-_{m,s}$,
\bea \delta^+_{m,p}\,g_{11}&=&\frac43(m+1)g_{11}
(w^1)^m(w^2)^p\,,\nn\\
\delta^+_{m,p}\,g_{12}&=&\Big(\frac{m+1}3g_{12}w^2
+p\,g_{11}w^1\Big)(w^1)^m(w^2)^{p-1}\,,\nn\\
\delta^+_{m,p}\,g_{22}&=&-2\Big(\frac{m+1}3g_{22}w^2
-p\,g_{12}w^1\Big)(w^1)^m(w^2)^{p-1}\,,\nn\\
\delta^+_{m,p}w^1&=&-(w^1)^{m+1}(w^2)^p\,,\quad
\delta^+_{m,p}\,w^2=0\,,\nn\\
&&-----------\nn\\
\delta^-_{m,p}\,g_{11}&=&-2\Big(\frac{p+1}3g_{11}
w^1-m g_{12}w^2\Big)(w^1)^{m-1}(w^2)^p\,,\nn\\
\delta^-_{m,p}\,g_{12}&=&\Big(\frac{p+1}3g_{12}w^1
+mg_{22}w^2\Big)(w^1)^{m-1}(w^2)^p\,,\nn\\
\delta^-_{m,p}\,g_{22}&=&\frac43(p+1)g_{22}
(w^1)^m(w^2)^p\,,\nn\\
\delta^-_{m,p}\,w^1&=&0\,,\quad\delta^-_{m,p}\, w^2=-(w^1)^m
(w^2)^{p+1}\,,\label{algebra.new.rep}\eea
where $m\in\mathbb{Z}$, $p\geq0$ for $\delta^+_{m,p}$ and
$p\geq-1$ for $\delta^-_{m,p}$. The range on $p$ is the minimal
truncation of (\ref{algebra.new.rep}) which includes all the
$SL(3,R)$ generators; but (\ref{covJ.def}) is satisfied by
arbitrary values of $m$ and $p$ for both operators
$\delta^\pm_{m,p}$. However, since all the fields are real while
$w^1$ and $w^2$ can be negative, the indices $m$ and $p$ only take
integer values.

The algebra satisfied by (\ref{algebra.new.rep}) is
\bea[\delta^+_{m,p}\,,\,\delta^+_{n,q}]&=&(m-n) \delta^+_{
m+n,p+q}\,,\quad[\delta^-_{m,p}\,,\,\delta^-_{n,q}]=(p-q)
\delta^-_{m+n,p+q}\,,\nn\\
~[\delta^+_{m,p}\,,\,\delta^-_{n,q}]&=&p\,\delta^+_{m+n,
p+q}-n\delta^-_{m+n,p+q}\,.\eea
Let $\delta^1_\bm =\delta^+_\bm$ and $\delta^2_\bm =\delta^-_\bm$,
with $\bm=(m^1,m^2)$ standing for the vector of indices, one cast
the above algebra into an even more compact form
\be[\delta^i_\bm\,,\,\delta^j_\bn]=m^j\delta^i_{\bm+\bn}
-n^i\delta^j_{\bm+\bn}\,,\label{algebra.new}\ee
where $i,j\in\{1,2\}$. This algebra is formally a type of
generalization to the usual Witt algebra. It has the general
structure of that for the map $\mathbb{R}'^n\mapsto\mathbb{R}'^n$,
(where $\mathbb{R}'$ means $\mathbb{R}$ with the points $0$ and
$\pm\infty$ pinched off,) for which $i,j\in\{1,\cdots,n\}$ and
$\bm=\{m^1,\cdots,m^n\}$. It will be interesting to see if
(\ref{algebra.new}) also appears in generic dimensions, for which
we expect $n=k-1$. (Loosely, one can firstly consider diagonally
embedding $SL(3,R)$ into $SL(4,R)$, then $SL(4,R)$ into $SL(5,R)$,
and so on; and (\ref{algebra.new}) is the most natural result to
expect.)

There is no fundamental obstacle to apply the same construction to
general situations. But for non-extremal black holes, the
subleading terms will make the calculation significantly more
involved. There is also the interesting question of what happens
if one starts with different embeddings of $SL(2,R)$ into
$SL(k,R)$. The choice (\ref{sl2r.g2}) is a diagonal embedding of
$SL(2,R)$ into $SL(k,R)$. We have also tried the principle
embedding of $SL(2,R)$ into $SL(3,R)$, but without success. (Our
preliminary result suggests that, for $SL(3,R)$, there is no
extension of the principally embedded $SL(2,R)$ to the Witt
algebra.) It should be interesting to find all possible algebras
that can arise from this construction in generic dimensions. We
note that the present construction is not sensitive to the
topology of horizons. So it gives the same result for the
Emparan-Reall black ring and the 5D Myers-Perry black hole.

Our result indicates that, in higher dimensions, the symmetry
governing the near horizon physics of a black hole could be much
larger than the previously known conformal symmetry. For black
holes with multiple independent rotating planes, it is known that
each non-vanishing rotation can give raise to a Virasoro algebra
and each is as good in reproducing the Bekenstein-Hawking entropy
through Cardy's formula \cite{Lu:2008jk}. Although there is a huge
difference between our construction and the calculations done
previously, all the methods must be related somehow if they are
going to describe the same physics. Our result then suggests that,
through similarity transformations, the different Virasoro's
corresponding to different rotations are in fact equivalent to
each other, and they are only part of a even larger symmetry such
as in (\ref{algebra.new}).

At the moment, an obvious task is to find a way to abstract
physical information from the newly found symmetries, which we
have not succeeded yet. Technically, this is partially due to the
fact that the symmetry found here is not the usual
diffeomorphisms, but is some internal gauge symmetry. As a result,
the successful techniques used in previous calculations (e.g.
those mentioned at the beginning of this paper) are not directly
applicable here. Finding a way out of this is a major goal of our
next step. Apart from helping us understand the properties of
black holes in better detail, it is also possible that the general
features found in generic dimensions can help us answer some of
the more direct questions in 4D.

\section*{Acknowledgement}

The author thanks Steven Carlip, Stefan Fredenhagen, Axel
Kleinschmidt, Stefan Theisen and especially Evgeny Skvortsov for
helpful discussions, comments and correspondence.

\end{document}